\documentclass[a4paper]{article}
\usepackage[T1]{fontenc} %
\usepackage{RR}
\usepackage{hyperref}
\usepackage{color}
\usepackage{cite}
\usepackage{graphicx}
\usepackage{algorithm}
\usepackage{algorithmic}
\usepackage{ntheorem}

\usepackage{amsmath,epsfig,color,url}
\usepackage{amsfonts}
\usepackage{graphicx}   
                                          
\usepackage{fancyhdr}
\usepackage{multirow}
\usepackage{mathrsfs}

\newcommand{\V}{\mathcal{V}}
\newcommand{\Orig}{\mathcal{O}}
\newcommand{\D}{\mathcal{D}}
\newcommand{\E}{\mathcal{E}}
\newcommand{\R}{\mathcal{R}}
\newcommand{\Rcoding}{\R_c}
\newcommand{\Rfwd}{\R_f}
\newcommand{\Nout}[1]{ \overrightarrow{ \mathcal{N}_{#1}} }
\newcommand{\Nin}[1]{ \overleftarrow{ \mathcal{N}_{#1}} }
\newcommand{\p}[2]{p_{{#1}{#2}}}
\newcommand{\Proba}{{\bf P}}
\newcommand{\Prob}[2]{\Proba_{{#1}{#2}}}
\newcommand{\auth}[1]{x_{#1}}
\newcommand{\code}[1]{c_{#1}}
\newcommand{\mode}[1]{m_{#1}}
\newcommand{\f}[1]{f_{#1}}
\newcommand{\Q}[1]{Q_{#1}}
\newcommand{\N}[1]{N_{#1}}
\newcommand{\fopt}{\mathcal{F}}
\newcommand{\sol}{\xi}

\newcommand{\attack}{\mathcal{A}}

\RRdate{March 2011}

\RRauthor{
Katia Jaffr\`es-Runser
  \thanks[sfn1]{\'Equipe SWING} 
  \and
	C\'edric Lauradoux
\thanksref{sfn1}
}
\authorhead{Jaffr\`es-Runser, Gorce \& Comaniciu}
\RRtitle{Planification pour la s\'ecurisation d'un codage en r\'eseau}
\RRetitle{Authentication planning for XOR network coding}
\titlehead{Authentication planning for XOR network coding}
\RRnote{This paper has been submitted to IEEE NetCod 2011}
\RRresume{
Les r\'eseaux sans-fil sont particuli\`erement vuln\'erables aux attaques par pollution dans lesquelles un attaquant externe est capable d'envoyer ces propres messages  sur le r\'eseau. Pour pouvoir d\'etecter de telles attaques  la destination, un code d'authentification (MAC: Message Authentication Code en anglais) est rajout\'e  chaque paquet. Un noeud interm\'ediaire peut v\'erifier la validit\'e d'un paquet de faon  limiter la port\'ee de transmission d'un paquet pollu\'e dans le r\'eseau.  
Dans le cadre d'un r\'eseau fortement contraint en \'energie tel qu'un r\'eseau de capteurs, le probl\`eme du d\'eploiement d'une strat\'egie de s\'ecurisation du r\'eseau
par MAC se pose. En effet, la consommation \'energ\'etique du r\'eseau sera fortement influenc\'ee d'une part par le type de MAC utilis\'e dans le r\'eseau et d'autre part par le choix des relais du r\'eseau qui v\'erifieront le code des paquets avant de les retransmettre. Nous nous int\'eressons plus particuli\`erement au cas o le r\'eseau de capteurs utilise une transmission par codage r\'eseau de par sa plus grande vuln\'erabilit\'e aux attaques par pollution. Ce type de r\'eseau n\'ecessite l'emploi de MAC d\'edi\'es (lin\'eaires).

Dans ces travaux, nous proposons une formulation combinatoire du probl\`eme de planification de la s\'ecurit\'e. Dans cette formulation, nous minimisons l'\'energie totale consomm\'ee par le r\'eseau s\'ecuris\'e pour la transmission d'un paquet par source dans le r\'eseau. Les variables sont les d\'ecisions d'authentification binaires des noeuds. Nous illustrons ce mod\`ele pour un r\'eseau papillon pour lequel diff\'erentes distributions des probabilit\'es d'attaque sur les liens sont consid\'er\'ees.
}
\RRabstract{
This paper formulates the authentication planning problem when network coding is implemented in a wireless sensor network. The planning problem aims at minimizing the energy consumed by the security application which is guarantied using message authentication codes. This paper proposes a binary non-linear optimization formulation for this planning problem whose decision variables are the authentication decision of the nodes and the MAC modes of operation. It is illustrated for a butterfly topology. Results show that there is a real trade-off between energy efficiency and message throughput in this context.         
}

\RRmotcle{R\'eseaux de capteurs, codage r\'eseau, s\'ecurit\'e, message authentication codes, planification, optimisation}
\RRkeyword{Wireless sensor networks, network coding, security, message authentication codes, planning, optimization}
\RRprojets{SWING}
\RRdomaine{1} 
\RRtheme{S\'ecurit\'e, R\'eseau, Codage en r\'eseau}
\URRhoneAlpes 
\RCGrenoble 

\begin{document}
\RRNo{7562}
\makeRR   

\section{Introduction}

Network coding~\cite{Ahlswede2000,Katti2008} are particularly vulnerable to \emph{pollution attacks}~\cite{Dong2009b} where an outsider adversary injects his malicious data. Indeed, network coding spreads the pollution by combining legitimate messages with polluted 
ones and therefore limiting the recovery probability of legitimate messages. Message authentication has to be ensured for  end-to-end communications between any source and destination of the network. Pollution attacks can be defeated using message authentication codes (MACs). The primary goal of MAC is to prevent an adversary to tamper with the messages (substitution) and to forge its own messages (impersonation). A keyed cryptographic digest of the message, which can be ciphered or not, is appended to the message. The message is authenticated successfully if the destination is able to compute the same keyed signature than the one appended to the message, knowing the secret key shared with the emitter. A comprehensive survey of MAC can be found in~\cite{Menezes1996}. In this paper, we make an extensive use of MAC based on universal hash functions (UHF-MAC)~\cite{Krawczyk1994}. Such functions may exhibit linearity which is particularly suited for network coding and they have been used in past works~\cite{Agrawal2009,Boneh2009,Apavatjrut2010} to thwart pollution attacks systematically by each node of the network.

In contrast to previous works~\cite{Agrawal2009,Boneh2009,Apavatjrut2010}, this paper addresses the problem of efficiently planning an authentication service for an energy constrained wireless network. A topical example is wireless sensor networking (WSN) whose security deployment has to guarantee low energy expenditure~\cite{Perrig2004}. It is used as a case study herein. The security planning  problem resumes to determining which nodes are going to authenticate the messages and which authentication strategies are the most energy efficient to deploy. We assume that the designer has some information on the distribution of the threat in the network. For instance, he may know that part of the network belongs to a trusted perimeter where security risks are low. Threat is modeled in this work with a probability of attack for each link of the network. A binary optimization formulation for the authentication planning problem is derived. Optimal solutions with respect to energy are provided and analyzed for a butterfly network topology with respect to various scenarios of attack.

The paper is structured as follows. Section \ref{sec:statement} states the problem and Section \ref{sec:model} derives the according optimization model. Section \ref{sec:results} gives energy optimal roll out strategies for the butterfly network 
and Section \ref{sec:conclu} concludes the paper.

\section{Authentication planning problem} \label{sec:statement}

In this paper, we only consider the case of XOR network coding~\cite{Katti2008} and not random linear network coding.

\begin{table*}
\caption{Energy performance of basic authentication strategies for a uniform attack topology.\label{tab:strategies}}
\begin{center}
\begin{tabular}{|c||c|c|c|}
\hline
					& Security maniac    		& Na\"ive strategy    				 		& Authentication planning\\ \hline \hline
Authenticating		& All relays +			& Destinations		 				& Optimal selection w.r.			\\  
 Nodes				& Destinations 			& only											&  threat + destinations			\\ \hline 
 Objective			& Detect threat asap		& Limit unnecessary checks  					& Minimize energy			 	\\	\hline 
Low threat 			& Energy wasted for		& Quasi energy 					 	& \multirow{4}{*}{Energy optimal} \\ 
(small $p$)			&  unnecessary checks	& optimal 						&  					\\ \cline{1-3}		 		
High threat 			& Quasi energy    		& Energy wasted for  					&							\\  
(high $p$)                   &  optimal   				&  forwarding polluted messages 						& \\ \hline
\end{tabular}
\end{center}
\end{table*}

\subsection{Attack topology}

Pollution attacks are committed on the links of the network $G$. The number of links attacked and their location define an attack topology $\attack$. We consider that the designer may not have a complete knowledge of the location of the attack at any point in time. Hence, he may have a confidence level in a link depending on its location in the network. For instance, links located inside of a trusted perimeter may have a higher confidence while the ones outside have a lower confidence. 
This feature is modeled using a probability of attack $\p{i}{j}$ on a link $(i,j)$ of the network which is defined as the probability of a message of being attacked on $(i,j)$. 

\begin{figure}[ht]
      \begin{center}
		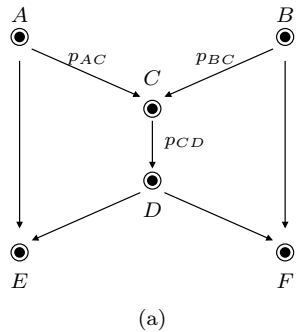
	\end{center}

   \caption{Butterfly network and attack topology.\label{fig:butt}}
\end{figure}

An attack topology $\attack$ is defined by the distribution of the probabilities of attack for all the links of the network (see Fig.~\ref{fig:butt}). For instance, this topology can be uniform and in this case, all links are attacked with the same probability $p$. Topology $\attack$ may as well model an attack localized on a single link $(i,j)$. In this case, $\attack$ is composed of a single non-null probability of attack $\p{i}{j}$. 
%

\subsection{Authentication strategies}

The purpose of this paper is to derive a model that yields the \emph{energy optimal authentication strategy} knowing an attack topology $\attack$ on a network $G$ performing XOR network coding. An authentication strategy is defined by the subset of nodes that authenticate the messages and the modes of UHF-MAC used (presented later in this section). 

First of all, it is important to note that the sources generate the messages with their corresponding digests and that the destinations always verify what they have received. Consequently, the destinations are always able to disregard polluted messages. These checks are mandatory and their cost in energy is incompressible.

For all relaying nodes other than the destinations, we have a degree of freedom: they may or not authenticate the messages. 
This may incur a certain authentication cost in energy at the relays. As shown in~\cite{Apavatjrut2010} performing verification is as energy expensive as sending plus receiving a message. However, in some cases it may be beneficial to the overall network energy performance since polluted messages are not uselessly forwarded towards the destination. In the case of network coding, it will also prevent the creation of polluted combination and preserve the throughput.
   
In terms of security planning, two extreme strategies are often considered. On the one side, the \emph{security maniac strategy} 
emphasizes on detecting an attack as soon as possible, limiting the pollution in the network. All the relaying nodes authenticate the messages as considered in~\cite{Agrawal2009,Boneh2009,Dong2009b}. Unnecessary verifications can lead to a waste of energy. On the other side, the \emph{security na\"ive strategy} considers that an end-to-end authentication is sufficient and that there is no need to empower relays with authentication capabilities. Forwarding polluted messages incurs both a same energy waste and a throughput reduction.

Intermediate strategies are possible to improve the energy and throughput performance. Table~\ref{tab:strategies} resumes all strategies. Finding optimal strategies is the aim of this work which can be achieved by solving an optimization problem. More specifically, we define the \emph{authentication planning problem} that minimizes the overall energy consumption of a WSN knowing its topology, the network coding rules, the  authentication MAC modes and the attack topology existing in the network.



\subsection{MAC schemes}
It has been established by Apavatjrut et al. in~\cite{Apavatjrut2010} that MACs based on the classical primitives that are block ciphers or hash functions imply an energetic cost too important for the relaying nodes of a WSN. The same observation holds for the underlying primitives (exponentiation) used in~\cite{Boneh2009}. On the opposite, MACs based on UHFs~\cite{Agrawal2009,Apavatjrut2010} offer more flexibility for the authentication if we use an $\epsilon$-almost XOR universal hash ($\epsilon$-AXU) function $h$ is~(see~\cite{Krawczyk1994} for more details). The most interesting property for our problem is the linearity of these functions: $h(m_1) \oplus h(m_2)=h(m_1 \oplus m_2)$ with $m_1$ and $m_2$ two $n$-bit messages. We voluntarily skip the details related to this function as they are not essential to understand the core of this paper (see ~\cite{Agrawal2009,Apavatjrut2010} for further details). 

Exploiting the linearity is particularly interesting for authentication in the context of network coding. Let us consider a node in the network who has to combine (XOR) $\ell$ messages and their corresponding authentication codes. A MAC based on $\epsilon$-AXU function offers three possibilities for authentication: (i) the node  authenticates each message individually, combines (XOR) the valid ones and computes the authentication code of this sum. Then, the message is forwarded. We refer to this mode of operation by AXF throughout the paper. The AXF mode requires $\ell$ verifications, \textit{i.e.} $\ell$ computations of the MAC. (ii) The node checks that the sum modulo two of the authentication codes is equal to the authentication codes of the sum modulo two of the messages. By doing so, it exploits the linearity of the MAC to reduce the authentication to a single computation of the MAC. 
We refer to this mode of operation as XAF. The drawback of the XAF is that the node forwards a message if and only if the $\ell$ incoming messages are not polluted. (iii) The node can also simply forward the sum modulo two of messages alongside the sum modulo two of the authentication codes. Any verification is delegated to other nodes. We refer to this mode of operation as XF.





\section{Optimization model}\label{sec:model}

The authentication planning problem is formulated in the following using a binary integer program. 

\subsection{Network model}

We assume that the network topology is known. The network is modeled using a directed acyclic graph $G(\V,\E)$ having vertex set $\V$ and edge set $\mathcal{\E \subset \V \times \V}$. 
Without loss of generality, $ \V = \{1, 2, \dots, |\V| \} $.
For each node $i \in \V$, $\Nout{i}$ and $\Nin{i}$ are the sets of edges leaving from and the set of edges going into $i$, respectively. 
Formally $\Nout{i} = \{ (i,j) | (i,j) \in \E\}$ and $\Nin{i} = \{ (j',i) | (j',i) \in \E\}$.

A set of sources $\Orig$ and destinations $\D$ is defined. Since we address the WSN case, $\Orig$ is the set of sensors having data to report in multicast to the $|\D|$ sink nodes and we consider $|\D| <<|\Orig|$. 
Source and destination nodes do not relay the information. As a consequence, the network we are modeling is composed of a set of relay nodes $\R = \V \setminus (\Orig \cup \D)$. In the following, we consider that the number of relays in the network is $N = |\R|$. 
So far, we do not consider any propagation losses and assume a perfect channel transmission.

The vertices $\V$ are partitioned into two groups of nodes: a subset $\Rcoding \in \R$ of relays performing XOR network coding and  $\Rfwd = \R \setminus \Rcoding$ which are simply forwarding messages. Knowing  $G(\V, \E)$, simple rules are set to define $\Rcoding$ and $\Rfwd$:  the coding relays have more than one edge coming into them (i.e. $|\Nin{i}{}|>1$) while forwarding-only relays $\Rfwd$ are characterized by a single incoming edge (i.e. $|\Nin{i}{}|=1$).

The binary quantity $\code{i} \in \{0,1\}$ is fixed to differentiate nodes of $\Rcoding$ performing network coding from nodes of $\Rfwd$ that are simply forwarding messages. Hence, for any node $i \in \Rcoding$, $\code{i}=1$ and for $i \in \Rfwd$, $\code{i}=0$.  

\paragraph*{Attack topology}
Security threats are modeled in $G$ by a valuation $\p{i}{j}$ on edge $(i,j) \in \E$. The set of all valuations $\attack = \lbrace \p i j | (i,j) \in \E \rbrace $ defines an attack topology. 
We assume that the attacks are independent on the edges of the graph. 
The attack topology is considered as being known by the network designer. 
In this setup, energy optimal security strategies can be derived as shown in the following. 

The following optimization model captures the impact of the previously described MAC strategies on the overall energy consumption of the network. 

\subsection{Optimization variables}


\paragraph*{\bf Authenticate variable}
Any node in $\R$ may or may not authenticate messages, whether this node is a coding or a forwarding-only node. 
Let $\auth{i} \in \{0,1\}$ be the first main binary variable of this model. If $\auth{i}=1$, node $i$ authenticates each incoming message while if $\auth{i}=0$, it never authenticates.
 
If a node authenticates a message (i.e. $\auth i = 1$), it has two opposite effects on the overall energy consumption:
\begin{itemize}
\item more energy is spent by the authentication process,
\item but energy for forwarding polluted messages is saved.
\end{itemize}

\paragraph*{\bf MAC mode of operation}
As shown earlier, XAF and AXF modes do not yield the same energy consumption. 
Let $\mode{i} \in \{0,1\}$ be a binary variable that gives the mode of authentication used by a coding node $i \in \Rcoding$. 
If $\mode{i}=1$, we have XAF and if $\mode{i}=0$ we have AXF. 
This variable can be interpreted as whether the XOR operation is done before authentication ($\mode{i}=1$ ; XAF mode) or after authentication ($\mode{i}=0$ ;  AXF mode). 

\paragraph*{\bf Node and network authentication strategy}
The \emph{node authentication strategy} $S_i$ is defined for any coding node $i \in \Rcoding$ by the tuple $S_i = (\auth i, \mode i, \code i)$ and for any forwarding-only node $i \in \Rfwd$ by the pair $S_i = (\auth i, \code i)$.  
Table \ref{tab:modes} gives the correspondence between the possible tuples and the MAC modes of operation for any coding and forwarding-only node. For forwarding-only nodes, the value of $\mode{i}$ is undetermined since no XOR step is performed. 

The \emph{network authentication strategy} $\sol$ is defined by the set of node authentication strategies for all nodes of the network $\sol = \lbrace S_i, \forall i \in \R \rbrace$.

\begin{table}[ht]
\caption{MAC modes and corresponding security strategies} \label{tab:modes}
\begin{center}
\begin{tabular}{|l|c|c|} \hline
\multicolumn{3}{|c|}{Node security strategies} \\ \hline
\multirow{3}{*}{Coding node} & AXF & $(\auth i=1,~\mode{i}=0,~ \code i =1)$ \\
& XAF & $(\auth i=1,~\mode{i}=1,~ \code i =1)$ \\
& XF &  $(\auth i=0,~\mode{i}=1,~ \code i =1)$  \\ \hline
\multirow{2}{*}{Relay node} & AF &  $(\auth i=1,~ \code i =0)$  \\
& F & $(\auth i=0,~ \code i =0)$\\ \hline
\end{tabular}
\end{center} 
\end{table}

\subsection{Forwarding decisions}

We define the\emph{ forwarding decision of a node $i$ as the probability that this node decides to transmit a received message}.  
Let $\f i \in \left[0,1\right]$ be the forwarding probability of node $i$. If authentication is performed by node $i$, this decision is positive if no polluted message is detected. This decision is a direct consequence of the node authentication strategy $S_i$ ($\auth i$, $\mode i$) and the probability of a polluted message to arrive from a direct neighbor node $k$ to node $i$. 

Let define $\Prob k i$ as the \emph{probability of a polluted message to arrive at node $i$ coming from node $k$}. This probability is a function of node authentication strategy $S_k$, the forwarding decisions and the attack probabilities related to all the paths between the sources $s \in \Orig$ and $i$ going through $k$. Its derivation is given after the forwarding decision description.   

The forwarding decision $f_i$ of node $i$ depends on its node type $\code i$.
The global formulation of $\f i$ is:
\begin{equation} 
\f i = (1-\code i)\cdot \f i ^{R} + \code i \cdot \f i ^{C} 
\label{eq:forwarding}
\end{equation}
where $\f i ^{R}$ and $\f i ^{C}$ are the forwarding probabilities for the case node $i$ is a forwarding or a coding node, respectively.

\paragraph{Forwarding node} For $i \in \Rfwd$ ($\code i = 0$), the forwarding decision is function of $\auth i$ and $\Prob k i$. 
If $a_i=0$, the node simply forwards every received message and hence $f_i=1$. Else ($\auth i=1$) the node forwards with probability $\f i = (1-\Prob k i )$ which represents the probability of an unpolluted message to arrive in $i$. The forwarding probability for the case $\code i = 0$ is derived as:
\begin{equation} 
 \f i ^{R} = (1-\auth i)+ \auth i  (1-\Prob k i) 
\end{equation}

\paragraph{Coding node} For $i \in \Rcoding$ ($\code i = 1$), the forwarding decision depends on the MAC strategy. For the case of AXF, a node forwards a message if at least one of its incoming messages is non-polluted which happens with probability $1-\prod_{k \in \Nin i} \Prob k i$.
For the case of XAF, a message is forwarded if all messages XOR-ed together are non-polluted which happens with probability $\prod_{k \in \Nin i} (1-\Prob k i)$

A closed-form derivation of the forwarding decision for any type of node of the network is given by:
\begin{equation} 
\begin{split}
\f i ^{C} = (1-\auth i)+ \\ \auth i \left[ 
	\mode i \cdot \prod_{k \in \Nin i} (1-\Prob k i) 
	+
	(1-\mode i)(1-\prod_{k \in \Nin i} \Prob k i)  
\right]
\end{split}
\end{equation}

\paragraph{The pollution probability $\Prob k i$} A message coming into node $i$ from a neighbor node can be polluted for two reasons: $i)$ node $k$ sends a message that is not polluted and the message gets polluted on the link between $k$ and $i$ following the local probability of attack $\p k i$ on $(k,i)~$ ; $~ii)$ node $k$ forwards a message that is polluted (this is only the case if node $k$ does not implement an authentication function, i.e. $\auth k = 0$). 

For the case $\auth k = 1$, node $k$ authenticates and the pollution probability is equal to $\Prob k i = \f k \cdot \p k i$, which is the probability that node $k$ forwards an unpolluted message and that it can only be polluted by an attack on link $(k,i)$.   

For the case $\auth k = 0$, node $k$ cannot detect if it forwards a polluted message or not. Hence, the probability for the message sent by $k$ to be polluted depends on the previous history of the message in the network. Hence, it is derived recursively knowing the values of the forwarding and attack probabilities on all paths coming into node $k$. In this case, $\Prob k i = 1-(1-\p k i) \cdot \prod_{l \in \Nin k} (1-\Prob l k)$, where $\prod_{l \in \Nin k} (1-\Prob l k)$ is the probability for a message to arrive in $k$ without being polluted on the links coming into $k$.

A global formulation of $\Prob k i$ with respect to $\auth k$ is given by: 
\begin{equation} \label{eq:prob}
\begin{split}
  \Prob k i = \f k \left[ 
  	\auth k \cdot \p k i  ~~+ \right. \\ \left.  
	(1-\auth k )\cdot
	\left( 
		1-(1-\p k i) \cdot \prod_{l \in \Nin k} (1-\Prob l k)
	\right) 
   \right] 
\end{split}
\end{equation}
For the case $k=s \in \Orig$, $\Prob s i = \p s i $ since $\f s = 1$.

\paragraph{Pollution and forwarding probability}

Since the network is a directed acyclic graph, there are no loops in the network and the values of the pollution and forwarding probabilities exist and can be derived for any node of the network. The causal dependency between the definitions of these probabilities is rooted in the network coding of the messages originating from different paths. In order to compute $\f i$ for node $i$, the pollution probabilities on all its incoming links $\{\Prob k i, \forall k \in \Nin i\}$ are needed. These values depend on the forwarding probabilities of the intermediary nodes that belong to the existing paths joining the source nodes to $i$.  
 
We consider first the case of a layered network where the network is divided into layers of nodes. The sources are connected to nodes of layer one but not to nodes of layer 2, nodes of layer 1 are connected to nodes of layer 2 but not to nodes of layer 3, etc. In this case, pollution and forwarding probabilities can be computed layer by layer. For layer 1 nodes, $\Prob s i = \p s i$ and $\f i$ is deduced using \eqref{eq:forwarding}. Then, $\Prob i j $ is computed according to \eqref{eq:prob} for layer 2 nodes and $\f j$ is derived according to \eqref{eq:forwarding} for layer 2 nodes as well. The process is repeated until the destination layer is reached. 
This iterative algorithm can be extended to support the case of a more general DAG, but for conciseness purposes, it is not presented herein.

\subsection{Energy cost function}

The energy cost function $\fopt_E(\sol)$ counts the energy spent for the end-to-end transmission of one message sent by the sources $s \in \Orig$ to their destinations $\D_s$ for a specific network authentication strategy (or solution) $\sol$:
\begin{equation} 
\fopt_E(\sol) = \fopt_\Orig(\sol) + \fopt_\R(\sol) + \fopt_\D(\sol)
\end{equation} 
where $\fopt_\Orig(\sol)$, $\fopt_\R(\sol)$ and $\fopt_\D(\sol)$ are the costs in energy relative to the energy expenditure of source nodes, relay nodes and destination nodes, respectively.

\begin{table}[ht]
\caption{Energy costs ($\times 10^{-4} J$) for the atomic actions. Values are given for the transmission of one message.} \label{tab:valsEneOptim}
\begin{center}
\begin{tabular} { |l|c|c|}
\hline
Emission & $Q_T$ & 0.556851 \\\hline
Reception & $Q_R$ & 0.7995405  \\ \hline
Authentication (UHF-MAC)& $Q_A$ &1.686154 \\ \hline	
XOR of 2 messages & $Q_{XOR}$ & 0.00003135 \\ \hline
\end{tabular}
\end{center} 
\end{table}

The costs in energy for the atomic actions are listed in Table \ref{tab:valsEneOptim}. They have been collected in \cite{Apavatjrut2010} using the WSim/eSimu energy estimation tool \cite{Fournel2009} for a TI MSP430 based platform and a Ti CC2420, 802.15.4 compliant, radio device similar to TelosB nodes.
It is worth mentioning that one authentication is as expensive as a combined message emission and reception. 

The cost related to the transmission of a message by the source nodes is directly proportional to the number of sources $\fopt_\Orig=|\Orig| \cdot (\Q T + \Q A)$.
The cost related to the reception of a message by the destination nodes depends on the number of messages $\N d$ destination $d$ will receive from its previous hop neighbors:
\begin{equation*}
\fopt_\D = \sum_{d \in \D} \N d \cdot (\Q R + \Q A) 
\end{equation*}
where $ \N i = \sum_{k \in \Nin i} \f k $.

The derivation of the energy consumption for all relays in the network is given by:
\begin{equation*}
\begin{split}
\fopt_\R = \sum_{i \in \R} \left[
		\N i \cdot \Q R + \auth i \cdot \fopt_A (\code i, \mode i) + \f i \cdot \Q T  
	\right]
\end{split}
\end{equation*}
The function  $\fopt_A (\code i, \mode i)$ gives the energy consumption of authentication with respect to the type of node (coding or forwarding) and the MAC mode considered. It is defined by:
\begin{equation*}
\begin{split}
  \fopt_A (\code i, \mode i)=  \code i \cdot  \left[ 
  		\Q {XOR} \cdot (\N i - 1) + \right. \\ \left.
		 \mode i \cdot \Q {XAF}(i) + (1-\mode i)\cdot \Q {AXF} (i) \right]   + (1-\code i) \Q A \cdot \Proba_i^{Rec}
\end{split}
\label{eq:foptA}
\end{equation*}
where $\Q {AXF}(i)$ and $\Q {XAF}(i)$ are the costs for authenticating a message using AXF and XAF, respectively. $\Proba_i^{Rec} = 1 - \prod_{k \in \Nin i} (1- \f k)$ is the probability of node $i$ to receive at least one message from its one hop neighbors.

XAF has the cost of authenticating a single message ($\Q {XAF} (i)= \Q A \cdot \Proba_i^{Rec}$) since it is performed on the XOR-ed version of the incoming messages. 
$\Q {AXF}(i)$ is a function of the number of messages received at node $i$ since each incoming message is authenticated individually. It is derived as $\Q {AXF} (i)= \Q A \cdot \N i$

\subsection{Optimization problem definition}

We recall that $\R$  is a set of $N$ relays of $c = |\Rcoding|$ coding and $N-c$ forwarding relays. 
The \emph{security planning problem} can be formulated by the binary integer program as follows.
 
\begin{eqnarray*}
	& \min 			&\fopt_E(\sol)\\
s.t. & \f i  =	 	 & (1-\code i)\cdot \f i ^{R} + \code i \cdot \f i ^{C} , ~~~~~				\forall i \in \R \\
	& \f i ^{R}=  	 & (1-\auth i)+ \auth i  (1-\Prob k i), ~~~~ 									\forall i \in \Rfwd \\
	& \f i ^{C}=  	 &(1-\auth i)+\auth i \left[ \mode i \cdot \prod_{k \in \Nin i} (1-\Prob k i) \right.+ \\ 
	&				 &\left.(1-\mode i)(1-\prod_{k \in \Nin i} \Prob k i)  \right], ~				\forall i \in \Rcoding \\
	&  \Prob k i = 	 & \f k \left[ \auth k \cdot \p k i  ~~+ \right. \\ 
	&				 & \left. (1-\auth k )\cdot \left(1-(1-\p k i) \cdot \prod_{l \in \Nin k} (1-\Prob l k) \right) \right] \\ 
	& \sol = 		& \lbrace S_i, \forall i \in \R \rbrace		\\
	& S_i = 		& \left\{	
							\begin{array}{lc} 
								(\auth i, \mode i, \code i)	\hspace{0.65in} & \forall i \in \Rcoding \\
								(\auth i,  \code i)	& \forall i \in \Rfwd \\
							\end{array}
						\right. \\ 
	& (\code i, \auth i) \in 	&\lbrace 0,1\rbrace \times \lbrace 0,1\rbrace 		\hspace{0.75in} \forall i \in \R	\\
	& \mode i  \in 	&\lbrace 0,1\rbrace 	 										\hspace{1.26in}\forall i \in \Rcoding\\
	& \p k i \in 		& \left[0, 1\right]												\hspace{1.3in}\forall (i,j) \in \E \\
\end{eqnarray*} 
The solution set has a cardinality of $3^c \cdot 2^{N-c}$. The energy cost is not linear and hence, the problem is not an binary linear program.
\begin{table}[h]
\caption{Description of the six solutions for the butterfly network} \label{tab:strategies}
\begin{center}
\begin{tabular} { |c|c|c|c|}
\hline
Strategy	 	& $(\auth C, \mode C)$									& $\auth D$				& Description \\ \hline
(XF ; F) 		& $(0, - )$ 						& $ 0$		& $C$, $D$ forward only \\ \hline
(XF ; AF)		& $(0, - )$ 						& $ 1$		& Authentication on $D$ only \\ \hline
(AXF ; F) 		& $(1, 0)$ 				& $ 0$		& AXF on $C$ only \\ \hline
(XAF ; F) 		& $(1, 1)$ 		& $ 0$		& XAF on $C$ only \\ \hline
(AXF ; AF) 	& $(1, 0)$ 		& $ 1$		& AXF on $C$, Auth. on $D$ only \\ \hline
(XAF ; F) 		& $(1, 1)$		& $ 1$		& XAF on $C$, Auth. on $D$ \\ \hline
\end{tabular}
\end{center} 
\end{table}

\section{Results}\label{sec:results}

\begin{figure}
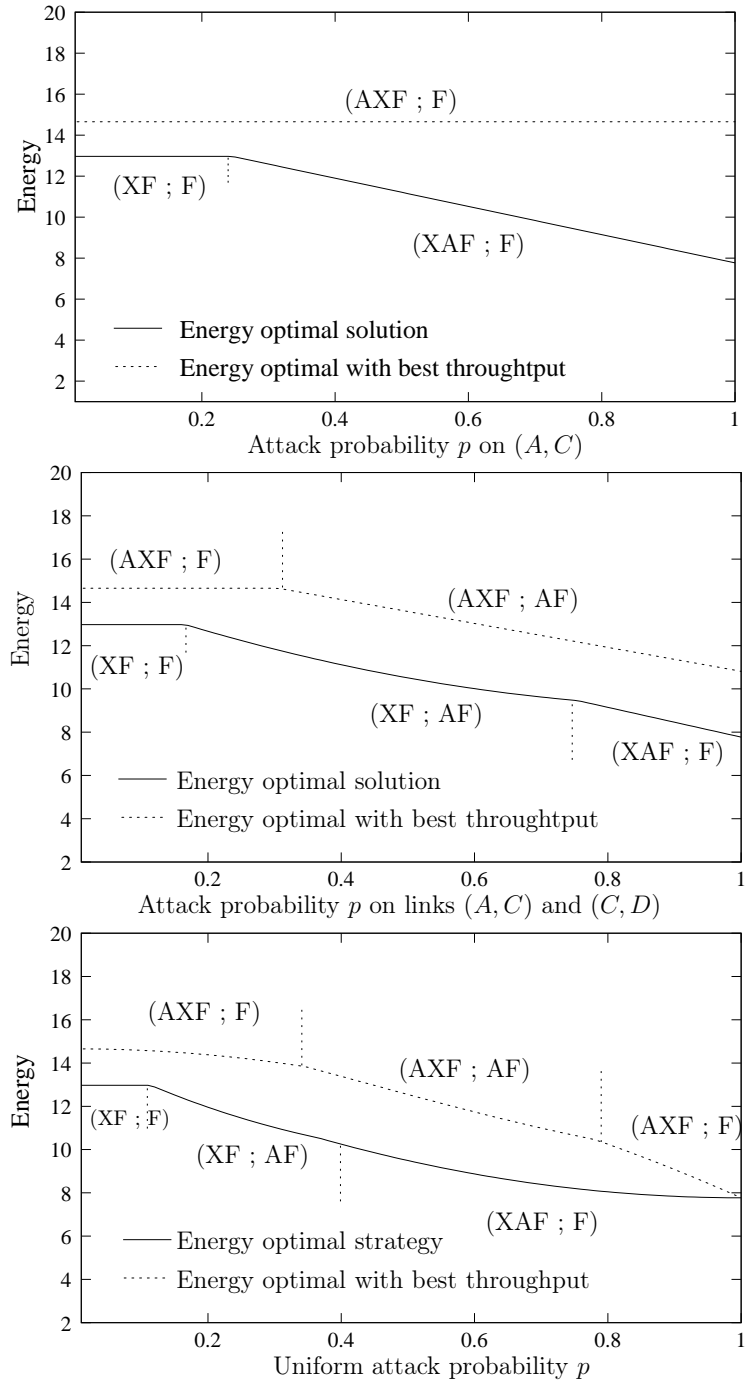

\begin{center}
\scalebox{0.8}{\input{fig/singleLinkAC.pstex_t}}
\scalebox{0.8}{\input{fig/twoLinksInARow.pstex_t}}
\scalebox{0.8}{\input{fig/uniform.pstex_t}}
\end{center}
\caption{Energy optimal strategy with and without best throughput constraint with respect to attack probability $p$ when 1, 2 and 3 links are attacked. \label{fig:resButt1}}
\end{figure}

\begin{figure*}
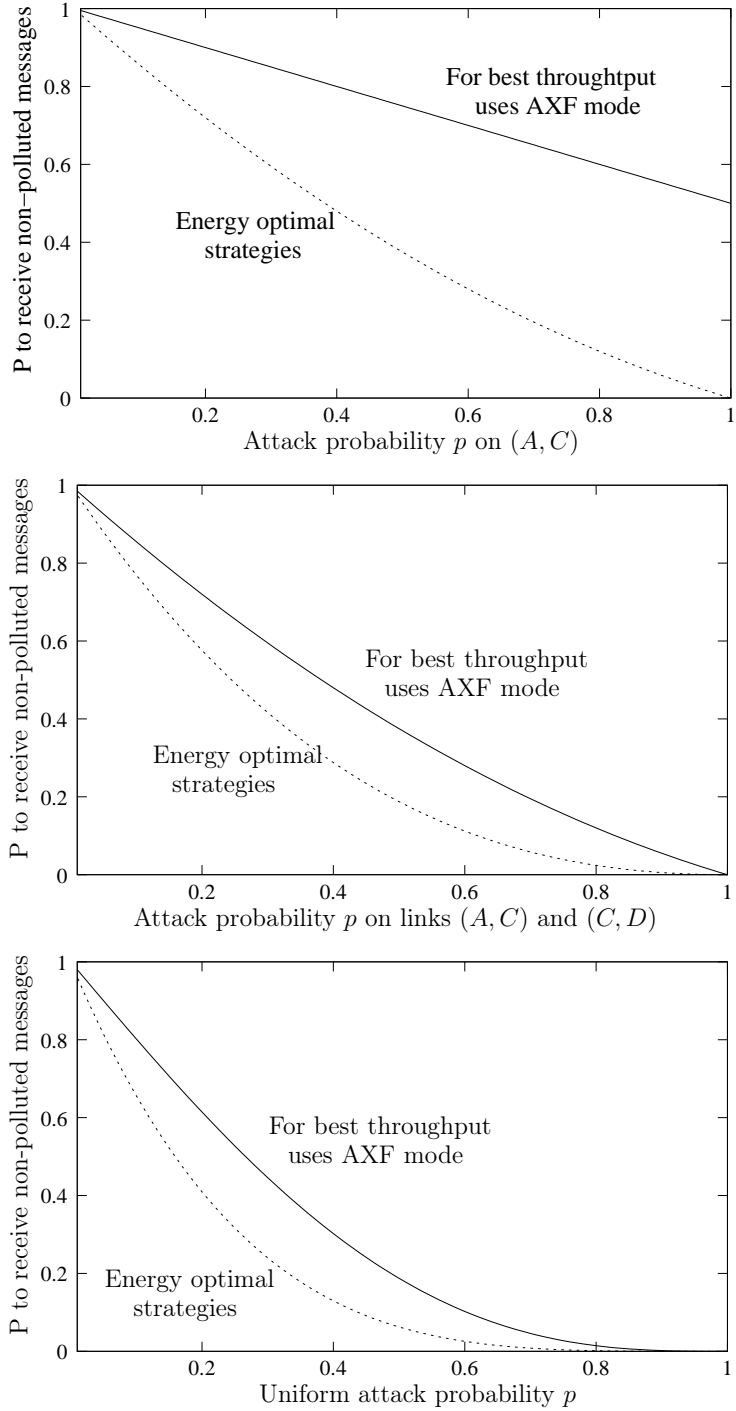

\begin{center}
\scalebox{0.8}{ \input{fig/singleLinkAC-P.pstex_t}}
\scalebox{0.8}{\input{fig/twoLinksInARow-P.pstex_t}}
\scalebox{0.8}{\input{fig/uniform-P.pstex_t}}
\end{center}
\caption{Average probability to receive non-polluted messages  for energy optimal strategies with and without best throughput constraint with respect to attack probability $p$ when 1, 2 and 3 links are attacked. \label{fig:resButt2}}
\end{figure*}

The security planning problem is illustrated for the butterfly network (cf. Fig.~\ref{fig:butt}).  
$C$ is a coding node and $D$ a forwarding relay. The six network authentication strategies are given in Table \ref{tab:strategies}. 
For instance, the first strategy (XF ; F) is the security na\"ive strategy where neither $C$ nor $D$ are authenticating. On the opposite, the two last entries are security maniac strategies where both nodes are authenticating.

In this analysis, we only consider attack topologies that involve the links $(A,C)$, $(B,C)$ and $(C,D)$ since attacks on other links can only be detected by destinations $E$ and $F$. 
We describe three scenarios of attack:
\begin{itemize}
\item The attack targets a \emph{single link}. Since the network is symmetrical, only the cases of an attack on links $(A,C)$ and $(C,D)$ are relevant. We chose to show the results for an attack on link $(A,C)$.  
\item The attack targets \emph{two links}. Again, only the cases where attacks are on the pair of links $(A,C)/(B,C)$ and $(A,C)/(C,D)$ can be considered for symmetry purposes. The results related to the $(A,C)/(C,D)$ pairs are presented here. We assume that the attacks on both links arise with the same probability (i.e. $p_{AC}=p_{CD}=p$).  
\item The attack targets the \emph{incoming links of the relaying nodes}. In this case, all three links are attacked with the same uniform probability (i.e. $p_{AC}=p_{BC}=p_{CD}=p)$.  
\end{itemize} 
The results related to the energy optimal strategies are given on Fig.~\ref{fig:resButt1} and Fig.~\ref{fig:resButt2}. Fig.~\ref{fig:resButt1} presents the total energy spent $\fopt_E(\sol^*)$ by the optimal strategy $\sol^*$ with respect to the attack probability $p$ on the link(s) targeted by the attacker. Two cases are considered. In the first one, we look for the \emph{energy optimal strategy} which minimizes total energy following the problem defined in Section \ref{sec:model}. In the second case, we show  the performance of the \emph{energy optimal with best throughput strategy} which looks for the strategy that maximizes the throughput at minimal energy.    

Throughput is measured in our case by the average probability $P_{th}$ for both destinations $E$ and $F$ to decode the messages of $A$ and $B$. Destinations can decode messages from $A$ and $B$ if they are non-polluted. If authentication is performed at coding node $C$, this probability depends on the MAC mode. For AXF, this probability is higher than for XAF because messages that are not polluted are always forwarded.  
A general expression for the butterfly network is $P_{th} = 0.5 \cdot f_c \cdot (1-p_{CD})\left[ 2 - p_{AC}-p_{BC}\right]$, where $f_C=1-p_{AC}\cdot p_{BC}$ for AXF and $f_C = (1-p_{AC})(1-p_{BC})$.  Fig.~\ref{fig:resButt2} gives the values of $P_{th}$ with respect to the attack probability $p$ for the \emph{energy optimal} and \emph{energy optimal with best throughput} strategies. 

Results show that strategies that minimize the number of forwarded messages are the most energy efficient ones. AXF never belongs to an \emph{energy optimal strategy} because the energy cost of verification is high but also because it transmits more messages. However, this is the only MAC mode that mitigates the spread of pollution induced by network coding. As a first conclusion, if throughput guarantee is the main concern, energy has to be spent for authentication by using AXF. 

When the probability of attack is low, the strategy where $C$ and $D$ simply forward messages is the most energy efficient since there are no security checks. However, when $p$ increases, more energy can be saved by authenticating messages at the relays. For instance, for a single attack on $(A,C)$ and $p > 0.24$, strategy (XAF;F) saves more energy because $D$ does not relay combined messages that are polluted. 
For high $p$, XAF mode on $C$ provides the minimum energy but drastically reduces throughput. Consequently, for small $p$, the \emph{energy optimal strategy} is to be favored because the loss of throughput is less important while for higher $p$, more energy-consuming modes of MAC have to be considered to favor throughput.  


\section{Conclusions and Perspectives }\label{sec:conclu}

This paper formulates the energy efficient authentication planning problem for XOR network coding. It formulates the problem as a binary non-linear optimization problem that minimizes the overall energy consumption of the secured network. Optimal roll-out of security-enabled nodes can be deduced together with their appropriate MAC mode. 
Results for the butterfly topology exhibit the trade-off between energy efficiency and throughput of non-polluted messages as a function of the MAC mode considered. 

Determining optimal security roll-outs should now be done for larger networks. In this context, exhaustive search is not scalable and proper optimization tools need to be derived in future works. A relaxed version of the problem can be formulated where the binary authentication variable becomes a message authentication probability. In this case, we move from a hard decision to a soft decision model, which could be easier to solve.
A multiobjective optimization approach could be considered as well in order to find the Pareto bound maximizing network throughput and minimizing energy expenditure. This work opens several perspectives as well. Extending this study to the case of random linear network coding is important for future work. Similarly, the definition of distributed algorithms that converge to energy efficient authentication strategies is a practical result of interest.

\bibliographystyle{IEEEtran}
\bibliography{IEEEabrv,mac}

\newpage

\tableofcontents

\end{document}